\pdfoutput=1
\documentclass[%
 reprint,
 superscriptaddress,
 amsmath,amssymb,
 aps,
 prl,
]{revtex4-2}

\usepackage{graphicx}% Include figure files
\usepackage{dcolumn}% Align table columns on decimal point
\usepackage{bm}% bold math
\usepackage{physics}
\usepackage{CJKutf8}
\usepackage{diagbox}
\begin{document}
%\linenumbers

\preprint{APS/123-QED}

\title{On-Demand Storage of Photonic Qubits at Telecom Wavelengths}

\author{Duan-Cheng Liu}
\author{Pei-Yun Li}
\author{Tian-Xiang Zhu}
\author{Liang Zheng}
\author{Jian-Yin Huang}
\author{Zong-Quan Zhou}
\email{email: zq\_zhou@ustc.edu.cn}
\author{Chuan-Feng Li}
\email{email: cfli@ustc.edu.cn}
\author{Guang-Can Guo}
%\affiliation{
% CAS Key Laboratory of Quantum Information, University of Science and\\
%Technology of China, Hefei, 230026, China
%}
%\affiliation{
% CAS Center For Excellence in Quantum Information and Quantum Physics,\\
%University of Science and Technology of China, Hefei, 230026, China
%}
\affiliation{
CAS Key Laboratory of Quantum Information, University of Science and Technology of China, Hefei, 230026, China
}
\affiliation{
CAS Center for Excellence in Quantum Information and Quantum Physics, University of Science and Technology of China, Hefei, 230026, China
}
\affiliation{
Hefei National Laboratory, University of Science and Technology of China, Hefei, 230088, China
}
\date{\today}

\begin{abstract}
Quantum memories at telecom wavelengths are crucial for the construction of large-scale quantum networks based on existing fiber networks. On-demand storage of telecom photonic qubits is an essential request for such networking applications but yet to be demonstrated. Here we demonstrate the storage and on-demand retrieval of telecom photonic qubits using a laser-written waveguide fabricated in an $^{167}$Er$^{3+}$:Y$_2$SiO$_5$ crystal. Both ends of the waveguide memory are directly connected with fiber arrays with a fiber-to-fiber efficiency of 51\%. Storage fidelity of 98.3(1)$\%$ can be obtained for time-bin qubits encoded with single-photon-level coherent pulses, which is far beyond the maximal fidelity that can be achieved with a classical measure and prepare strategy. This device features high reliability, easy scalability and can be directly integrated into fiber networks, which could play an essential role in fiber-based quantum networks.
\end{abstract}

\maketitle

%\section{Introduction} 

%\textit{Introduction.}--
Due to the inevitable photon loss, long-distance quantum communication is a challenging task in fiber networks. A quantum repeater can solve this problem based on quantum memories and entanglement swapping \cite{Sangouard2011}. An elementary link of a quantum repeater, i.e., heralded distribution of two-party entanglements, has been demonstrated using various systems, such as diamond defects \cite{hensen2015loophole}, trapped ions \cite{moehring2007entanglement}, single atoms \cite{hofmann2012heralded}, quantum dots \cite{delteil2016generation} and atomic ensembles \cite{yu2020entanglement,liu2021heralded}. 
Nevertheless, these systems are not able to provide direct telecom interfaces for fiber networks. 

Er$^{3+}$ in solids has highly coherent optical and spin transitions \cite{bottger2009effects,ranvcic2018coherence} and is naturally compatible with the telecom C-band which allows long-distance transmission of photons. Significant progress has been made for telecom-wavelength quantum memories using the atomic frequency comb (AFC) protocol \cite{jin2015telecom,saglamyurek2015quantum,saglamyurek2016multiplexed,askarani2019storage,craiciu2019nanophotonic,PhysRevResearch.3.L032054} albeit with predetermined storage times. Storage of classical telecom light with dynamic control can be realized in an on-chip memory device \cite{craiciu2021multifunctional}. On-demand storage of weak coherent pulses has been achieved using controlled reversible inhomogeneous broadening protocol \cite{lauritzen2010telecommunication}, while the low achieved efficiency and the resulting low signal-to-noise ratio are insufficient to support storage of qubits. Other than Er$^{3+}$-doped solids, telecom photons can also be linked with quantum memories by operating with nondegenerate photon pair source \cite{lago2021telecom,PhysRevLett.127.210502}, quantum frequency conversion \cite{PhysRevLett.123.063601,PhysRevLett.124.010510,van2022entangling,bock2018high,yu2020entanglement}, or mechanical resonators \cite{wallucks2020quantum}.
On-demand storage of photonic qubits from external sources is essential for the construction of efficient quantum repeaters based on absorptive quantum memories \cite{liu2021heralded,tang2015storage,Sangouard2011} but yet to be demonstrated at the telecom wavelengths. 

Here we demonstrate a fiber-integrated quantum memory at telecom wavelengths based on a laser-written waveguide fabricated in an $^{167}$Er$^{3+}$:Y$_2$SiO$_5$ crystal. On-demand storage of photonic qubits is achieved with Stark modulation of the AFC storage process using on-chip electrodes \cite{horvath2021noise,liu2020demand,craiciu2021multifunctional}. The experimental set-up is illustrated in Fig. \ref{fig:setup}. 

The substrate is an $^{167}$Er$^{3+}$:Y$_2$SiO$_5$ crystal. Y$_2$SiO$_5$ is a widely employed host material due to the low magnetic moments of its constituent elements (Y). $^{167}$Er is the only stable erbium isotope with nonzero nuclear spin which provides long ground-state population and coherence lifetimes, given that the Er$^{3+}$ electronic spin flips have been inhibited with ultra-low temperatures and strong magnetic fields \cite{bottger2009effects,ranvcic2018coherence,PhysRevB.105.245134}. To facilitate easy operations, an integrated device with direct fiber connection is preferred for such applications. Among various micro-/nano fabrication techniques aimed at integrated quantum memory, femtosecond-laser micromachining (FLM) introduces low damage to substrates and allows three-dimensional fabrication. The storage efficiency of integrated quantum memories in Er-doped crystals are currently limited to $0.2 \%$ \cite{askarani2019storage,craiciu2019nanophotonic}, while telecom-wavelength quantum memory has yet to be demonstrated in FLM waveguides.

Here we fabricated a type-III waveguide \cite{chen2014optical} along the D2 axis in an $^{167}$Er$^{3+}$:Y$_2$SiO$_5$ crystal. The host crystal has a doping level of 50 ppm of $^{167}$Er$^{3+}$.
The symmetric, circular-shaped cross section of a type-III waveguide can support light guidance of arbitrary polarization and facilitates efficient integration with single-mode optical fibers. 
The fabrication parameters are optimized according to the fiber-coupling efficiency, as discussed in Sec. I of the Supplemental Material \cite{supplementary}.

\begin{figure*}[t]
%\begin{center}
\includegraphics [width= 2.0\columnwidth]{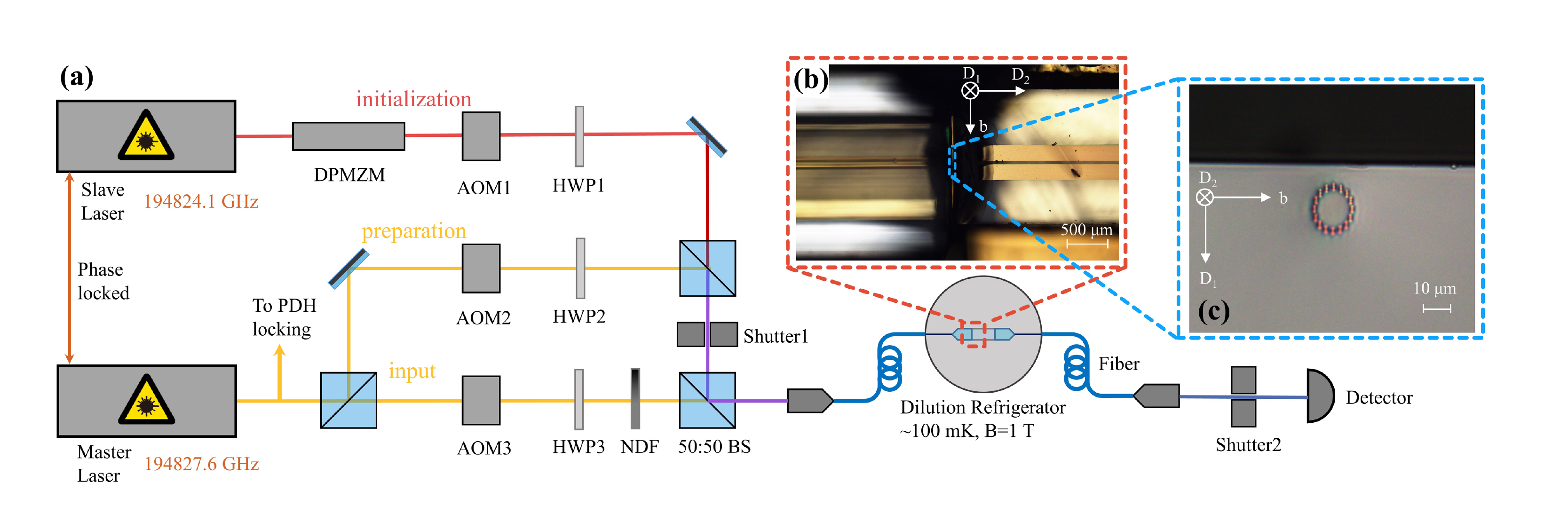}
\caption{\label{fig:setup} Diagram of the experimental setup. (a) The optical path before cryostat can be divided into three sections: one for spectral initialization, one for AFC preparation, and one for input pulses.
All beams are controlled by acousto-optic modulators (AOM) in double-pass configurations. A dual-parallel Mach-Zehnder modulator (DPMZM) is employed for wideband sweeping during the spectral initialization. The input pulse is attenuated to weak coherent state by neutral density filters (NDF). All beams are combined with two 50:50 beam splitters (BS) and collected into a single-mode fiber connecting to the memory. Polarization of the laser of each beam is adjusted with half wave plates (HWP) to optimize the sample absorption. Two mechanical shutters are employed to protect the single-photon detector from strong laser. 
 (b) A micrograph of the memory device. On the left is a single-channel fiber array. On the right is the $^{167}$Er$^{3+}$:Y$_2$SiO$_5$ crystal with a laser-written waveguide inside (too thin to be visible) and gold electrodes on top of it. (c) The cross section of the laser-written waveguide.}
%\end{center}
\end{figure*}

Single-mode fibers (SMFs) were pigtailed at both ends of the waveguide. The contact between the fiber and waveguide is solidified with UV glue. To make the coupling more stable, as shown in Fig. \ref{fig:setup}(b), the fibers are encapsulated into V-groove assemblies as fiber arrays. A sketch of fiber-to-waveguide connection is shown in Fig. S1 in the Supplemental Material \cite{supplementary}. The fiber-to-fiber efficiency of this device is 51$\%$, indicating a good matching between the waveguide mode and the fiber mode. Such device efficiency has been comparable to that of commercial fiber-integrated devices such as electro-optic modulators. After mounting into the cryostat, the total efficiency drops to 25$\%$ due to additional losses caused by four fiber connectors which can be avoided by fusion splicing of the fibers in future works. A detailed discussion of the total transmission efficiency is given in Sec. V of the Supplemental Material \cite{supplementary}.

A pair of gold strip electrodes were placed near both sides of the optical waveguide to apply electric field pulses for the implementation of Stark-modulated AFC storage protocol. Details on the fabrication process are given in Sec. II of the Supplemental Material \cite{supplementary}.

Due to the large unquenched electronic angular momenta, Er$^{3+}$ ions in solids are prone to experience spin flips and flip-flops, which will lead to spectral diffusion and short ground-state population lifetimes \cite{bottger2006optical,baldit2010identification}, undermining the performances in the storage time and the efficiency. The flip-flops can be reduced when the electronic spins are polarized with a sufficiently low temperature and strong magnetic field \cite{PhysRevB.105.245134,ranvcic2018coherence}. Electronic spin flips can be reduced if ultralow temperature is used to directly reduce the phonon density \cite{PhysRevB.105.245134}, or when a sufficiently strong magnetic field is applied to make most of the phonons off-resonant with the electronic Zeeman transition \cite{ranvcic2018coherence}. In this work Er$^{3+}$ ions in site II \cite{bottger2006spectroscopy} are addressed, and the magnetic field is applied along the D1 axis of the host crystal for a large electronic Zeeman energy splitting \cite{guillot2006hyperfine}. Here the lowest working temperature is approximately 100 mK and the magnetic field is 1 T, which is sufficient to create the spectral feature required in this experiment.
A higher working temperature can be tolerated if a stronger magnetic field is employed \cite{PhysRevResearch.3.L032054}.

\begin{figure*}[t]
%\begin{center}
\includegraphics [width=2.0\columnwidth]{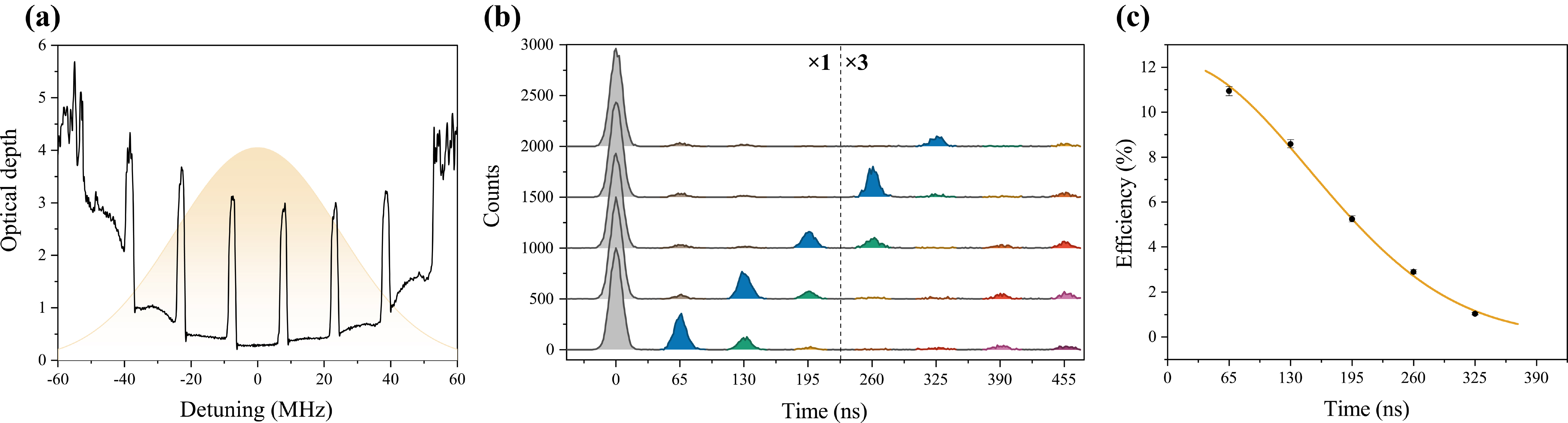}
\caption{\label{fig:echo} On-demand storage of single-photon-level inputs. (a) The AFC structure (black line) with a comb spacing of 15.4 MHz and an illustration of the spectral distribution of input pulses (shown in yellow). (b) Photon-counting histogram of on-demand storage with input pulses containing 0.4 photons per pulse on average. 500 counts are shifted along y-axis between each set of data. The transmitted input pulses are shown in grey. The on-demand readout echoes are shown in blue. The subsequent emissions are given in other colors. The data after 230 ns is magnified by 3 times for visual convenience. (c) The storage efficiency as a function of readout time. The yellow line is a fit based on Eq. (S2). Here $ \gamma $ is fitted to be 1.8 MHz, which is approximately consistent with the measured AFC structure in Fig. \ref{fig:echo}(a).}
%\end{center}
\end{figure*}

Our photonic memory is based on the Stark-modulated AFC protocol \cite{horvath2021noise,craiciu2021multifunctional,liu2020demand}. AFC is a periodic spectral profile of an inhomogeneously-broadened atomic ensemble \cite{afzelius2009multimode}. After the AFC absorbs a single photon the quantum state of the atom ensemble undergoes phase evolution according to the transition frequencies of individual atoms. This results in a pre-determined photon re-emission at the time of $n/\Delta$ where $\Delta$ denotes the periodicity of AFC and n represents an arbitrary positive integer. Such phase evolution can be actively controlled by modulating atomic transition frequencies through the linear Stark effect, leading to on-demand retrieval of the signal in discrete steps \cite{horvath2021noise,liu2020demand,craiciu2021multifunctional}. Details on the Stark-modulated AFC protocol and the Stark coefficient are provided in Sec. III of the Supplemental Material \cite{supplementary}.

Spectral initialization was performed before AFC preparation to enhance the absorption depth and the storage efficiency. As illustrated in Fig. S3 and detailed Sec. IV of the Supplemental Material \cite{supplementary}, optical pumping was performed roughly in the middle part of the $\Delta m_I=0$ absorption band to enhance its side parts. Here $m_I$ denotes the quantum number of the nuclear spin. Our electromagnet can provide a maximal magnetic field of up to 1 T and in this case the transitions of $\Delta m_I=\pm1$ and $\Delta m_I=0$ were not completely resolved in frequency so that the state initialization was performed partially \cite{ranvcic2018coherence, craiciu2019nanophotonic}. Nevertheless, the spectral initialization process helped to enhance the sample absorption by a magnitude of more than two times, which was crucial for high-efficiency storage.

Fig. \ref{fig:echo}(a) presents an example of the prepared AFC which has a bandwidth of 100 MHz and $\Delta$ of 15.4 MHz. Fig. \ref{fig:echo}(b) shows the on-demand retrieval of AFC echoes controlled by applying two electric pulses at different times. These electric pulses have a TTL-compatible voltage of $\pm$4.825 V with  the generated electric field of $\pm$1.2 kV/cm. The pulse duration is set as 18 ns. The input weak coherent state pulse is Gaussian-shaped with a FWHM pulse width of 15 ns and, on average, contains 0.4 photon per pulse. Time dependence of the storage efficiency is presented in Fig. \ref{fig:echo}(c). The retrieval efficiency is calculated as the AFC echo counts divided by the input pulse counts. The efficiency is $10.9\%\pm0.2\%$ at a storage time of 65 ns. Compared to previous demonstrations of integrated absorptive telecom quantum memories (including fiber-based ones) \cite{jin2015telecom,saglamyurek2015quantum,saglamyurek2016multiplexed,askarani2019storage,craiciu2019nanophotonic}, the storage efficiency has been enhanced by more than 5 times. Due to the low damages introduced by the FLM fabrication and the hyperfine initialization process, the efficiency here is close the best result reported in bulk crystals \cite{ PhysRevResearch.3.L032054}. The available absorption depth still set a limit on the achievable storage efficiency, which can be further enhanced through a complete state initialization \cite{ranvcic2018coherence, PhysRevResearch.3.L032054} and longer crystals. Storage efficiency towards unity can be obtained by incorporating an impedance-matched optical cavity \cite{sabooni2013efficient}. 

\begin{figure}[p]
%\begin{center}
\includegraphics [width= 1.0 \columnwidth]{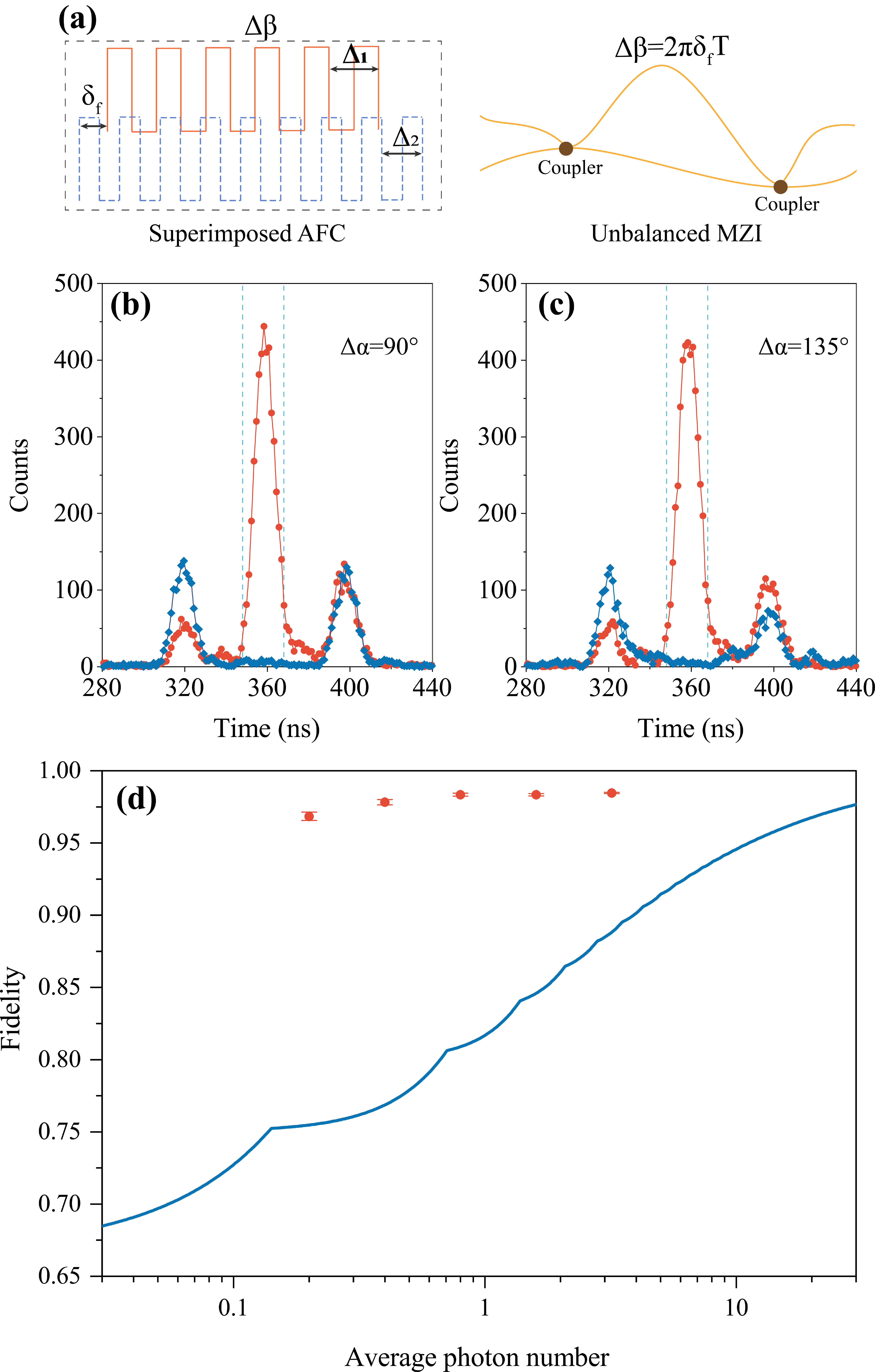}
\caption{\label{fig:timebin} On-demand storage of time-bin qubits. (a) A pair of superimposed AFCs serves as an unbalanced Mach-Zehnder interferometer (MZI) with adjustable relative phase between two paths. A frequency shift of $\delta_f$ for one AFC introduces a phase shift of $\Delta\beta=2\pi {\delta}_f T$ where $T$ is the relevant storage time (320 ns). (b) and (c) present the photon-counting histograms for the constructive (shown in red) and destructive (shown in blue) interference for input states of $\ket{e}+i\ket{l}$ and $\ket{e}+e^{i3\pi/4}\ket{l}$, respectively. Here the average input photon number $\mu=0.8$. Raw visibility of $95.6\%\pm0.4\%$ and $96.0\%\pm0.4\%$ are obtained, respectively. More photon counting histograms are provided in Fig. S6 and Fig. S7 in the Supplemental Material \cite{supplementary}. (d) Total storage fidelity versus the average input photon number per qubit. The red points are experimental results with error bars representing one standard deviations. The blue solid line shows the classical bound which takes into account the finite storage efficiency (6.9$\%$ for 320 ns) and the Poissonian statistics of the input fields \cite{gundougan2015solid,ma2021elimination,specht2011single}. Details on the analysis of storage fidelity are provided in Sec. VII of the Supplemental Material \cite{supplementary}.}
%\end{center}
\end{figure}

%\textit{On-demand storage of time-bin qubits.}--
To rigorously benchmark the quantum storage performances of this device, we measured the fidelity for storage of time-bin qubits, which are particularly robust in long-distance fiber transmission. Time-bin qubit state of $\ket{e}+e^{i\Delta\alpha}\ket{l}$ was generated with an AOM with a controlled phase $\Delta\alpha$. Each qubit was composed of two 12-ns pulses with a time interval of 40 ns. Four input states of $\ket{e}$, $\ket{l}$, $\ket{e}+i\ket{l}$ and $\ket{e}+e^{i\frac{3}{4}\pi}\ket{l}$ were used to characterize the storage process with varied average input photon number $\mu$ of 0.2, 0.4, 0.8, 1.6 and 3.2. Electric control was configured to on-demand retrieve the signal as the second-order AFC echoes. Here AFC with longer $1/\Delta$ storage times was employed for the storage of time-bin qubits with low cross-talks. Details about the AFC structures employed here are provided in Fig. S5 and Fig. S6 in the Supplemental Material \cite{supplementary}. 

For $\ket{e}$ and $\ket{l}$, the storage time is 320 ns with an efficiency of $6.9\%\pm0.1\%$. We obtain average fidelities $F_{el}$ from 98.3$\%$ to 99.7$\%$ with increased input photon numbers. For superposition states $\ket{e}+i\ket{l}$ and $\ket{e}+e^{i\frac{3}{4}\pi}\ket{l}$, one will need an unbalanced Mach-Zehnder interferometer to project on superposition states. Here we employed a double-AFC architecture \cite{liu2020demand}, which served as a built-in Mach-Zehnder interferometer, to analyze the superposition states. As shown in Fig. \ref{fig:timebin}(a), two AFC structures are superimposed with the $1/\Delta$ storage times of 160 ns and 180 ns, respectively. The relative phase ($\Delta\beta$) between the two paths of such interferometer can be adjusted by shifting the center frequency of one AFC. Projections onto arbitrary superposition states can then be performed. Fig. \ref{fig:timebin}(b) and (c) show the interference results for input states of $ \ket{e}+i\ket{l} $ and $ \ket{e}+e^{i3\pi/4}\ket{l} $ with $\mu=0.8$. Here the fidelity is $F_+ =97.8\%\pm0.2\%$ and $F_- =98.0\%\pm0.2\%$ for $ \ket{e}+i\ket{l} $ and $ \ket{e}+e^{i3\pi/4}\ket{l}$, respectively. The average fidelity $ F_{+-}=(F_+ +F_- )/2 $ is $97.9\pm0.2\%$. 

The total storage fidelity, as defined by $ F_T=\frac{1}{3}F_{el}+\frac{2}{3}F_{+-} $ \cite{marcikic2003long,gundougan2015solid,ma2021elimination}, is $98.3\%\pm0.1\%$ for $\mu=0.8$. This result violates the the maximal fidelity (80.9$\%$) that can be achieved by a classical measure-and-prepare strategy \cite{gundougan2015solid,ma2021elimination,specht2011single} by 174 standard deviations, unambiguously demonstrating the reliability of this quantum storage device. Fidelities of various average photon numbers are listed in Table S1 in Supplemental Material \cite{supplementary} and compared with the classical bound in Fig. \ref{fig:timebin}(d). 
The fidelities with all input levels are far beyond the classical bound. For the lowest input level in our measurements ($\mu=0.2$), the obtained $F_T$ violates the classical bound by 71 standard deviations. The storage fidelity is not sensitive to the input photon number because of the low noise floor. The unconditional noise probability is $(2.4\pm0.5)\times10^{-4}$ at the detection window. The signal-to-noise ratio is $72\pm 11$ when $\mu =0.2$. The remaining imperfections in fidelity are caused by imperfect measurements, such as the imbalanced storage efficiencies for $\ket{e}$ and $\ket{l}$ with double-AFC, and the inaccuracy in the control of $\Delta\beta$.

%\section{Discussion.}

%\textit{Discussion.}--
In conclusion, on-demand storage of telecom photonic qubits is demonstrated in a fiber-integrated quantum memory with high fidelity. Both the fiber arrays and FLM waveguides allow for three-dimensional configurations, which would enable high-density spatial multiplexing for efficient quantum repeaters \cite{liu2021heralded,Sangouard2011}. Substantially longer optical coherence lifetime of $^{167}$Er$^{3+}$:Y$_2$SiO$_5$ can be obtained with larger magnetic fields along preferred orientations where the spectral diffusion effect becomes lower \cite{bottger2009effects, horvath2021noise}, so that to enable applications in non-hierarchical quantum repeater architectures \cite{sinclair2014spectral,muralidharan2016optimal}. For more flexible applications, much longer storage times on the second scale \cite{ranvcic2018coherence} could be obtained with spin-wave storage protocols such as spin-wave AFC \cite{gundougan2015solid} or the noiseless photon echo protocol \cite{ma2021elimination}. Our waveguide structure strongly confines the light field so that much larger Rabi frequencies can be obtained \cite{liu2020reliable} to apply the optical $\pi$ pulses required for spin-wave memories \cite{PhysRevResearch.3.L032054}. The close-to-surface optical waveguides can further enable efficient interface with on-chip electric waveguides to facilitate spin decoherence control through dynamical decoupling. With all these developments, the Er-based material can play an essential role in large-scale quantum networks, as it already did in classical fiber networks.

\begin{acknowledgments}
%\section{Acknowledgments}
This work is supported by the National Key R\&D Program of China (No. 2017YFA0304100), Innovation Program for Quantum Science and Technology (No. 2021ZD0301200), the National Natural Science Foundation of China (Nos. 12222411 and 11821404) and this work was partially carried out at the USTC Center for Micro and Nanoscale Research and Fabrication. Z.-Q.Z acknowledges the support from the Youth Innovation Promotion Association CAS.

Note added.—During the preparation of this manuscript, we became aware of a related experiment that demonstrates telecom heralded quantum storage in a fiber-integrated laser-written waveguide \cite{doi:10.1126/sciadv.abn3919}.

D.-C.L. and P.-Y.L. contributed equally to this work.

\end{acknowledgments}


%\nocite{*}

% Produces the bibliography via BibTeX.
%\bibliography{main}
%

%\nocite{*}

% Produces the bibliography via BibTeX.
%\renewcommand\thefigure{\Alph{section}\arabic{figure}}    
\setcounter{figure}{0}
\renewcommand*{\thefigure}{S\arabic{figure}}
\renewcommand*{\thetable}{S\Roman{table}}
\renewcommand*{\theequation}{S\arabic{equation}}

\onecolumngrid
\maketitle
\section*{Supplemental Material}
\subsection{I. Fabrication of the type-III optical waveguide}
The type-III waveguide was fabricated in an $^{167}$Er$^{3+}$:Y$_2$SiO$_5$ crystal that was cut along its D1, D2 and b axes with dimensions of 2$\times$9$\times$2 mm$^3$ by a femtosecond-laser micromachining (FLM) system (WOPhotonics). The femtosecond laser has a wavelength of 1030 nm and the pulse duration is 210 fs. The repetition rate of the laser pulses is 201.9 kHz. We employed low-energy laser pulses to fabricate the waveguide to avoid damage to the crystal and to maintain the physical properties of the Er$^{3+}$ ions inside the waveguide. This waveguide has a diameter of 12 $\mu m$ and its center is located 20 $\mu m$ below the D2-b surface. The border of the type-III waveguide is formed by 16 etched lines patterned along the D2 axis. Different pulse energies are required to etch the border lines at different depths. For convenience, here the top half and bottom half of the waveguide were fabricated using pulse energies of 58 nJ and 62 nJ, respectively. The laser pulse is focused by a 100× objective with a numerical aperture of 0.7. The moving speed of the laser beam is 1 mm/s along the D2-axis of the crystal. All of these fabricating parameters were optimized according to the fiber coupling efficiency. Inside the waveguide, the 1/e optical 
coherence lifetime of Er$^{3+}$ ions was measured to be approximately 700 $\mu s$ at 1 T with the lowest working temperature. 

\begin{figure*}
%\begin{center}
\includegraphics [width= 0.8 \columnwidth]{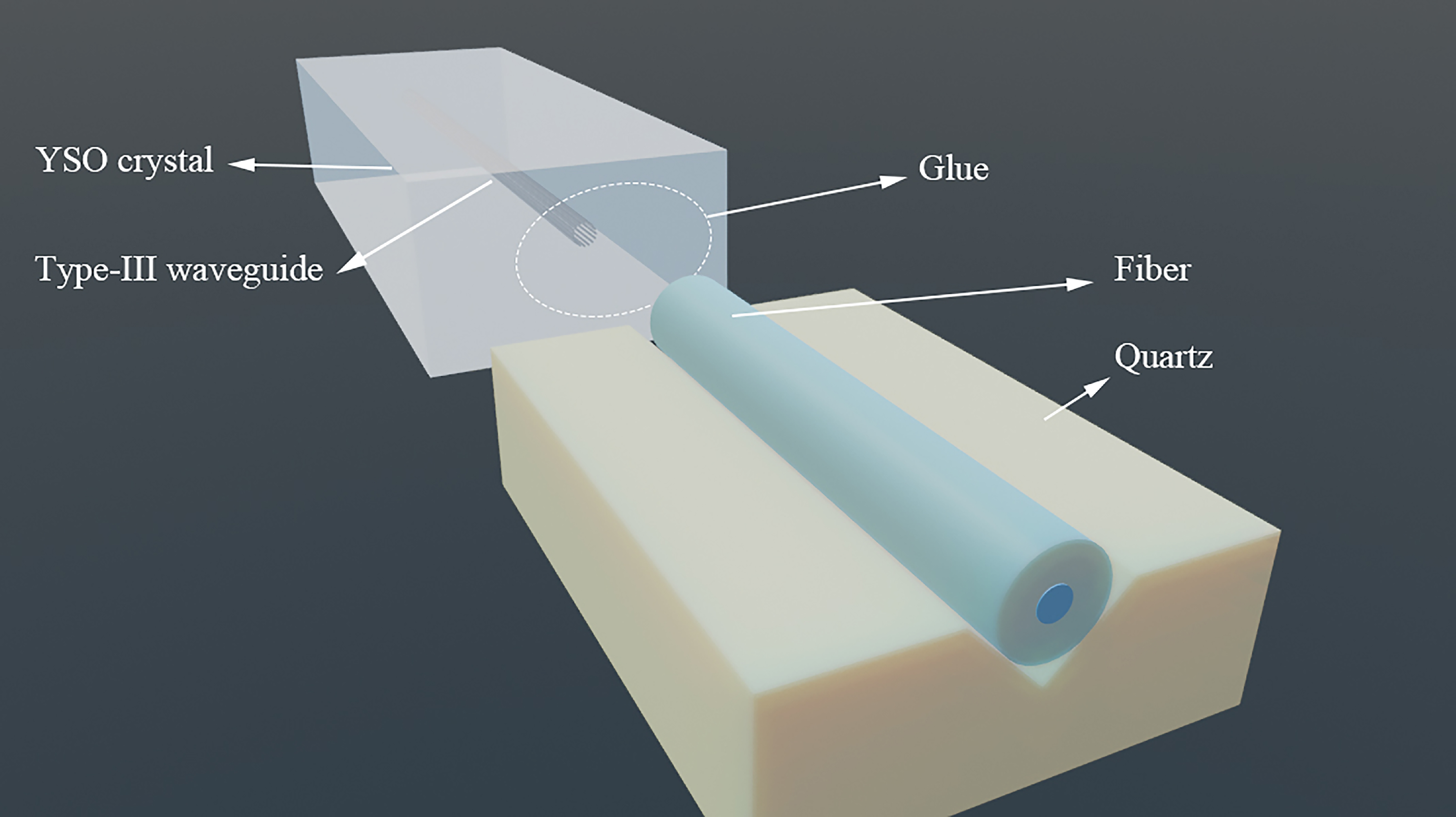}
\caption{\label{fig:array}A sketch of waveguide-fiber connectorization. For clarity, the sizes of the waveguide and the core of the optical fiber are exaggerated, and the lid of the fiber array is omitted.}
%\end{center}
\end{figure*}

\subsection{II. Fabrication of the electrodes}
Each of the electrodes had a width of 200 $\mu m$, and the spacing between the two electrodes was 40 $\mu m$. Electrodes were patterned onto the host crystal before the FLM fabrication process using standard UV lithography, e-beam coating and lift-off techniques. The electrode was consists of three layers. The bottom layer was 30 nm of titanium to enhance the adhesion between the crystal and the metal. The middle layer was 750 nm of copper. Finally, 20 nm of gold covered the copper to prevent oxidation. The speed of electron beam evaporation was 1 nm/s.

\subsection{III. The Stark-modulated AFC protocol}
After absorbing a single photon, the atomic ensemble can be described as a Dicke state \cite{afzelius2009multimode}: 
\begin{equation} \label{eq1}
|\Psi (t)\rangle =\Sigma_j e^{-i2\pi \delta_j t}e^{-ikz_j}|g_1...e_j...g_N...\rangle,
\end{equation}
Here $z_j$ represents the position of each atom and 
$\delta _j=m_j \Delta$ ($m_j =\pm 1, \pm 2, \pm 3 ...$)
is the frequency detuning between the input photon and the j-th atom of the AFC. k is the wave vector of input light and $|g_j \rangle$ ($|e_j \rangle$) represents the ground state (excited state) of an atom. This collective state will dephase rapidly because of the different phase factor of  $e^{-i2\pi \delta _j t}$ experienced by each atom in the comb. Nevertheless, rephasing comes at every $t=\frac{n}{\Delta }$(n=1, 2, 3...), and the echoes will be released. The decaying profile with respect to the $n$th-order echoes is dependent on the characteristics of the individual peaks in the comb. This can be observed from Eq. (A6) and Eq. (A18) in Ref. \cite{afzelius2009multimode}. There could be some collapse and revival when the comb profiles are not strictly Gaussian, as displayed in Fig. 2 in the main text.

The theoretical storage efficiency for Stark-modulated AFC is \cite{afzelius2009multimode,horvath2021noise}:  
\begin{equation} \label{eq2}
\eta = \widetilde{d}^2e^{-\widetilde{d}}e^{-\widetilde{\gamma}^2t^2}e^{-d_0},
\end{equation}
where $d_0$ represents the background absorption. $ \widetilde{d}=\frac{d}{F}\sqrt{\frac{\pi}{4\ln 2 }}$ denotes the effective absorption depth of the AFC where $F=\gamma/\Delta$ is the finesse of the AFC. $ \widetilde{\gamma} $ is related to the full width at half maximum (FWHM) of each absorption peaks $\gamma$ by $\widetilde{\gamma}=2\pi\gamma/\sqrt{8\ln 2}$ assuming a Gaussian lineshape. According to Eq. (S2), long memory lifetimes can be realized with an AFC with a higher finesse although the storage efficiency is sacrificed due to a reduced effective absorption depth. As an example, the storage performances for an AFC with a finesse of $14.5\pm0.3$ are given in Fig. S5(a).

Due to the symmetric properties of Y$_2$SiO$_5$, the electric field applied along crystal's b axis splits the transition frequencies of Er$^{3+}$ in the two subsites with the same magnitude while in opposite directions \cite{horvath2021noise}. When an electric field is applied, the optical transition frequencies of the two sub-classes of ions will shift by $\pm \Omega$, respectively. The ensemble state can be rewritten as: 

\begin{equation} \label{eq3}
|\Psi (t)\rangle =\Sigma_j e^{-i2\pi (\delta_j \pm \Omega) t}e^{-ikz_j}|g_1...e_j...g_N...\rangle,
\end{equation}

One electric pulse induces an extra phase shift of $e^{i2\pi \Omega T_p}$ for one sub-class and $e^{-i2\pi \Omega T_p}$ for the other. Here $T_p$ is the pulse duration of the electric field, while $\Omega=-\vec{\delta\mu}\cdot\hat{L}\cdot\vec{E}/\hbar$ denotes the magnitude of frequency shifts. $\vec{\delta\mu}$ represents the difference between the ground and excited-state electric dipole moments; $\vec{E}$ is the applied electric field and $ \hat{L}$ is the local field correction tensor \cite{macfarlane2007optical}. 
In case when an electric pulse with $\Omega T_p=1/4$ is applied during the time window (0, $1/\Delta$), two sub-classes of Er$^{3+}$ accumulate an extra relative phase shift of $\pi$. The successive AFC re-emissions are therefore silenced by destructive interference. To recover photon re-emissions, a second electric pulse with the same pulse length and magnitude while in the opposite direction can be applied during the time ($(n-1)/\Delta$, $n/\Delta$) and the echo will be finally readout at $t=n/\Delta$. 
%The extra phase will prevent the rephasing of the ensemble, and silent the echoes. When an opposite electric field with the same duration is applied, the extra phase can be compensated, and the subsequent AFC echoes can re-emit again.

\begin{figure*}
%\begin{center}
\includegraphics [width= 0.8 \columnwidth]{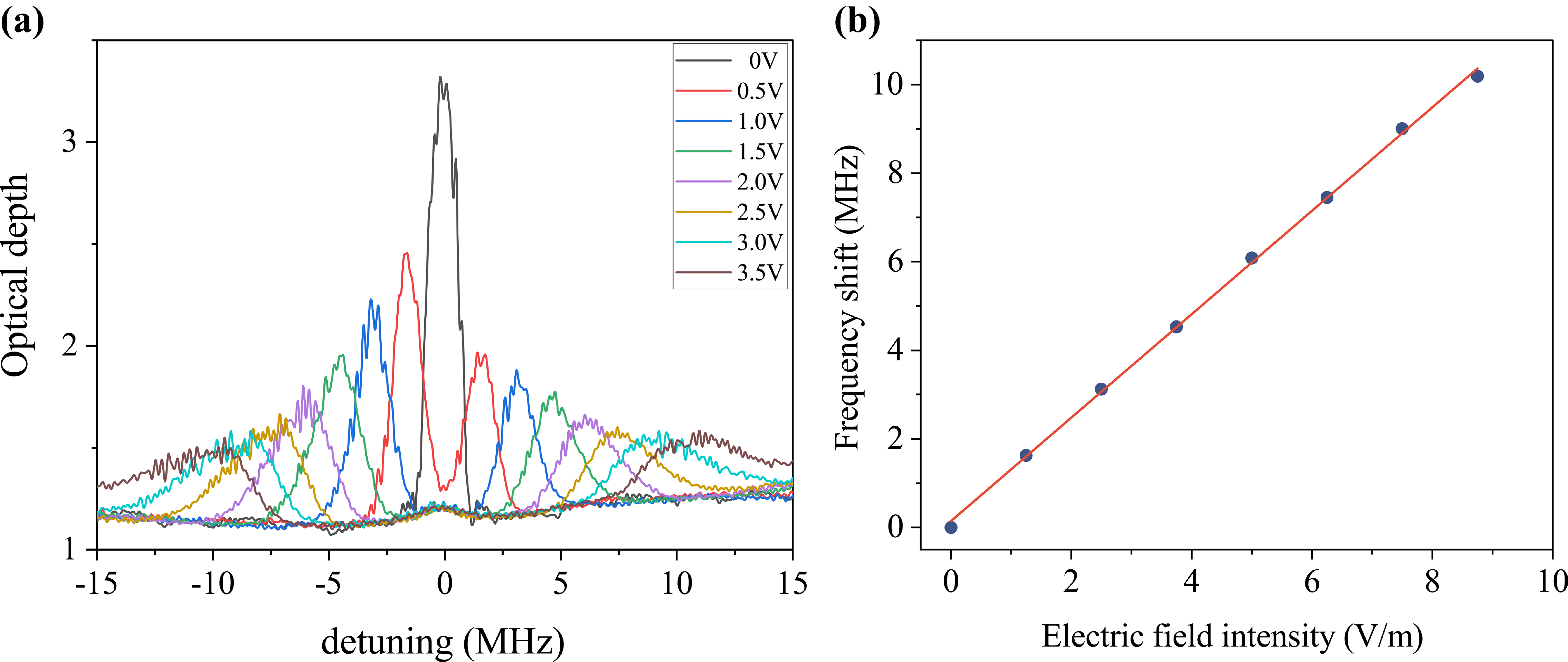}
\caption{\label{fig:stark} Measurements of the Stark coefficient. (a) The absorption peak is split by the electric fields. The bandwidths of the absorption peaks are broadened by the electric fields, indicating the inhomogeneity of the field distribution. (b) Frequency shift of one of the absorption peaks as a function of the electric field intensity. The red line is a linear fit. The stark coefficient along the b-axis of Y$ _2$SiO$_5 $ crystal is fitted as 11.68$\pm$0.24 kHz/(V/cm).
}
%\end{center}
\end{figure*}

Without the knowledge of $\vec{\delta\mu}$, DC stark effect was experimentally characterized with spectral-hole-burning spectroscopy \cite{liu2020demand}. An absorption peak was created inside a transparent spectral window, and split linearly along with the applied DC field along the b axis (Fig. \ref{fig:stark}). The Stark coefficient was measured as 11.68$\pm$0.24 kHz/(V/cm), which is consistent with the result reported in Ref. \cite{craiciu2021multifunctional}.

\subsection{IV. Optical Set-up and Spectral Initialization}
As shown in the Fig. 1(a), the master laser was locked at 194824.1 GHz with a stabilized Fabry-Perot cavity. This laser was employed to prepare the AFC and generate the input pulses. Another laser was phase locked to the first laser, with a working frequency of 194827.6 GHz. A dual-parallel Mach-Zehnder modulator (DPMZM) controlled with a vector signal generator (R\&S, SMW200A) was employed for single-sideband frequency sweeping to achieve spectral initialization and enhance the absorption depth. The fiber integrated memory device is cooled within a dilution refrigerator integrated with an external magnet. Signals at the single-photon level were analyzed by an InGaAs single-photon detector (ID230, IDQuantique) with a quantum efficiency of 25$\%$.

To our knowledge, the value of the transition dipole moment of Er$^{3+}$ in Y$_2$SiO$_5$ is unknown. In Y$_2$SiO$_5$ crystals, Er$^{3+}$ ions occupy crystallographic sites with very low local symmetry of C$_1$. This makes the transition dipole moment highly anisotropic, and its principle axes not necessarily parallel to the crystal’s optical extinction axes. With the impact of birefringence, it is hard to maintain the projection of the light polarization to a specific principle axis of the transition dipole moment.
Nevertheless, the input polarization in this experiment is optimized according to the measured absorption before performing the spectral initialization process. 

Spectral initialization is performed to further enhance the sample absorption for high-efficiency quantum storage. At a magnetic field of 1 T, the three absorption peaks of $^{167}$Er$^{3+}$:Y$_2$SiO$_5$ corresponding to $\Delta m_I=0$ and $\Delta m_I=\pm1$ \cite{ranvcic2018coherence} were not completely separated and the state initialization can only be applied partially \cite{craiciu2019nanophotonic}. The initialization was achieved by sweeping the laser with a bandwidth of 800 MHz whose center frequency was red shifted 500 MHz from the center of the absorption peak. 

\begin{figure*}
%\begin{center}
\includegraphics [width= 0.8 \columnwidth]{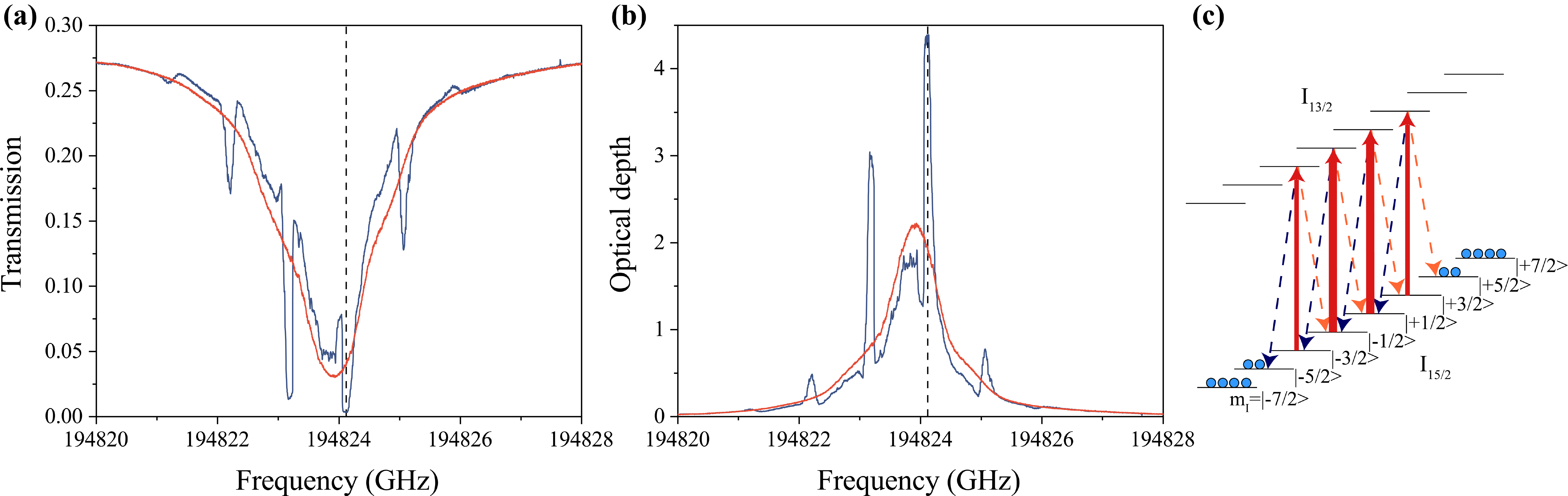}
\caption{\label{fig:ini} Spectral Initialization. (a) and (b) show the transmission and the optical depth of the $^{167}$Er$^{3+}$:Y$_2$SiO$_5$ waveguide. The red lines and the blue lines correspond to the conditions before and after the spectral initialization process, respectively. The dotted line denotes the center of enhanced absorption which is used to prepare AFC. (c) A sketch of the spectral initialization process. Here the blue circles represent the population transferred among the hyperfine states. Optical pumping is represented by red arrows. Orange and blue dashed arrows represent spontaneous decay with $\Delta m_{I}>0$ and $\Delta m_{I}<0$, respectively. }
%\end{center}
\end{figure*}

The $\Delta m_I =0$ absorption band can be considered as a superposition of inhomogeneously broadened absorption lines corresponding to all the eight $m_I$. Each of the absorption line has a distinct center frequency, while those locating on the sides tend to have large $|m_I |$. Therefore, when the middle part of the $\Delta m_I =0$ absorption band is pumped, population tend to transfer to hyperfine states with large $|m_I |$, as demonstrated in Fig. S3(c). Meanwhile, at a relatively lower magnetic field of 1 T, $\Delta m_I =0$ transitions corresponding to high $m_I$ values (+7/2, +5/2, ...) and $\Delta m_I =\pm 1$ transitions corresponding to low $m_I$ values (-7/2, -5/2, ...) tend to overlap. As a result, absorption at the dotted-line position in Fig. \ref{fig:ini} will be substantially enhanced. 

The frequency sweeping was performed with single-sideband modulation using a DPMZM (MXIQER-LN-30, iXBlue). Typically when optical signal is sinusoidally modulated by an electro-optic modulator (EOM) with the carrier frequency of $f_0$ and a modulation frequency of $\omega$, mutiple sidebands with frequencies of $f_0\pm n\omega$ ($n=0,1,2$...) will be generated. Here the DPMZM consists of two sub-Mach-Zehnder interferometers (MZIs) nested inside a third one. In this work both the two sub-MZIs were biased to their minimal point ($\pi$), while the third one was biased to -$\pi/2$. With this configuration the sideband of $f_0-\omega$ is maintained, while disturbing terms, including the carrier $f_0$ itself, as well as the sidebands of $f_0+\omega$ and $f_0\pm2\omega$, are cancelled due to phase mismatching \cite{izutsu1981integrated}. Fig. \ref{fig:ini}(a) and \ref{fig:ini}(b) show the changes of optical transmission and absorption depth before and after the optical pumping. The optical depth was enhanced by more than two times of magnitude through the initialization process. 

\subsection{V. The total transmission efficiency when mounting inside the cryostat}
In this work the device efficiency of 25\% was measured when the system cooled down while the device efficiency of 51\% was measured before this device was mounted into the cryostat. Independent measurements of the mating sleeves were performed at room temperature. The vacuum fiber feedthrough we used has a relatively low transmission efficiency of 80\%. Mating sleeves inside the cryostat have an efficiency of approximately 93\%. Therefore the overall transmission efficiency becomes $51\%×(80\%)^2×(93\%)^2=28.2\%$. This additional loss ($\sim$3.2\%) may come from fiber bending inside the cryostat and some small distortion during the cool-down process may also have contributions.

\subsection{VI. Time Sequences}
\begin{figure*}[h]
%\begin{center}
\includegraphics [width= 0.8 \columnwidth]{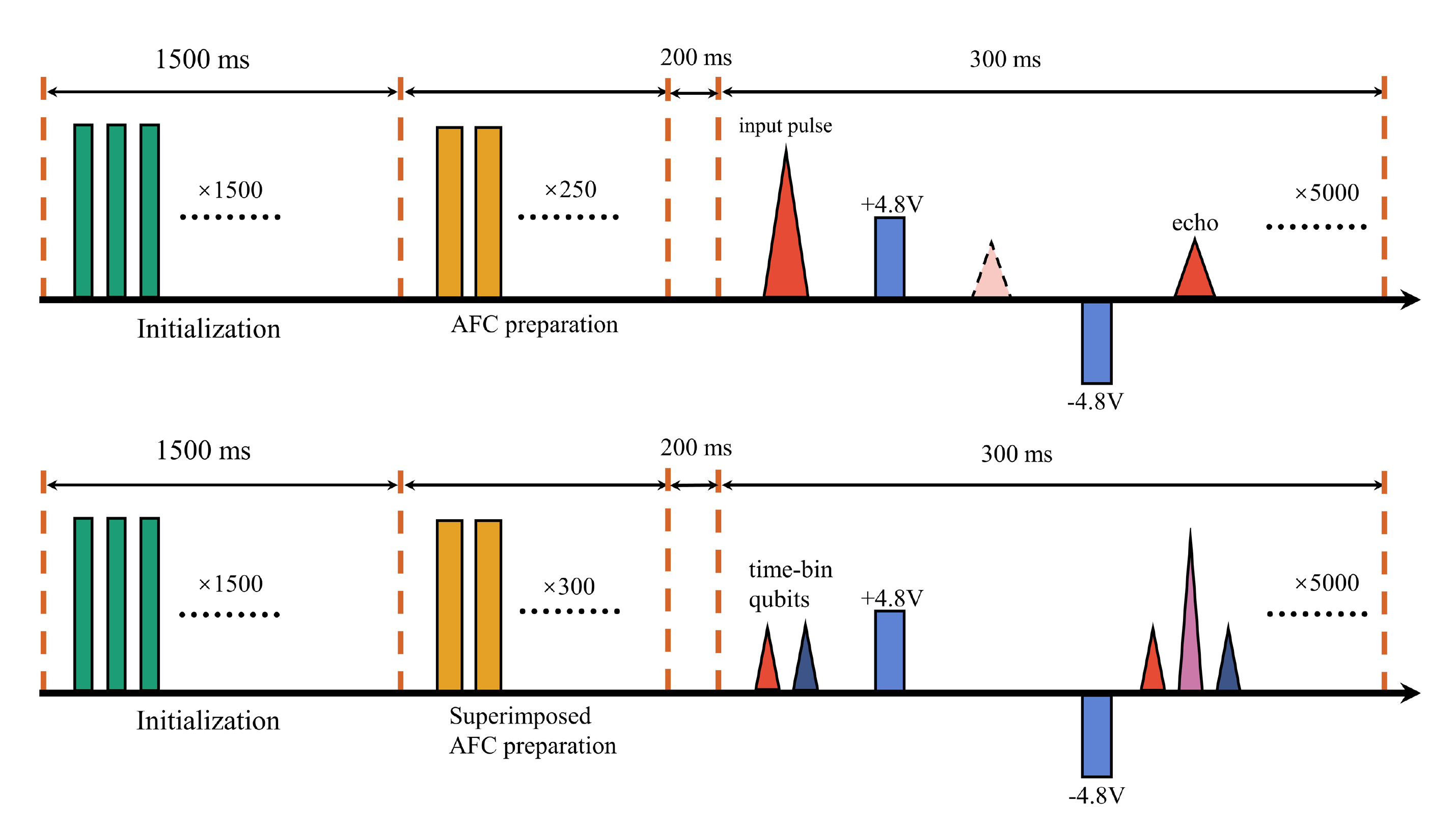}
\caption{\label{fig:sequence} Time sequence for the experiments. (a) The time sequence for on-demand storage of single-photon-level coherent pulses. (b) The time sequence for on-demand storage of time-bin qubits. }
%\end{center}
\end{figure*}

The complete time sequence of this experiment, including initialization, AFC preparation and storage process, is presented in Fig. \ref{fig:sequence}. During the initialization process, the pump pulse was repeated for 1500 times. Each pump pulse was chirped by 800 MHz during a time interval of 1 ms. For single AFCs, the AFC preparation was performed in a parallel manner \cite{jobez2016towards}. The AFC preparation pulse had a length of 5 ms, and 
was repeated for 250 times. For superimposed AFCs, 30 spectral pits were created one-by-one. 50-$\mu$s preparation pulses were used for each of the pits and these pulses were repeated by 300 times. The storage process started at 200 ms after AFC preparation, in order to avoid the influence of spontaneous emission. The input pulse was repeated for 5000 times with periodicity of 50 $\mu$s. 

\subsection{VII. The assessment of qubit storage fidelity}

For storage of time-bin qubits, the $1/\Delta$ storage time is extended to 160 ns. The efficiency decay curve of this AFC is presented in Fig. \ref{fig:echo}(c) and (d). The measured AFC structure is provided in Fig. \ref{fig:afc}(a). Photon-counting histograms for storage of basis states $\ket{e}$ and $\ket{l}$ are presented in Fig. \ref{fig:afc}(b). The input pulse had a full width at half maximum of 12 ns with average input photon number $\mu=0.8$. The fidelity of $\ket{e}$ and $\ket{l}$ is $99.1\%\pm0.1\%$ and $99.3\%\pm0.1\%$, respectively. Here the integration window was set as 20 ns while a narrower window could lead to a higher storage fidelity. The fidelity $F_e(F_l)$ is expressed as $(S+N)/(S+2N)$ where $N$ is the noise counts and $S$ represents the signal with noise floor excluded. The average fidelity is defined as
$F_{el}=(F_e +F_l )/2$.

For analysis of superposition time-bin qubits, double-AFC with $1/\Delta$ storage times of 160 ns and 180 ns were employed. The AFC structures with various $\Delta\beta$ that were used in measurements of time-bin qubits are shown in Fig. \ref{fig:afc}(c-f). %The total fidelity for storage of time-bin qubits can be defined as $ F_T=\frac{1}{3}F_{el}+\frac{2}{3}F_{+-} $\cite{afcsingle}. The storage performances for superposition states have been described in the main text. 
The fidelity of a superposition state is evaluated from the interference visibility $ V=(max-min)/(max+min) $ by $ F=(1+V)/2 $. Here $max$ and $min$ were measured in the cases when $ \Delta\beta = -\Delta\alpha $ and $ \Delta\beta = -\Delta\alpha+\pi $, respectively.
Fig. \ref{fig:el} shows the measurement of time-bin qubits with average photon number of 0.2. The experimentally determined fidelity and the maximal fidelity that can be obtained with a classical measure-and-prepare strategy are compared in Fig. 3 in the main text. The classical fidelity bound is derived according to Eq. 4 from the Supplementary Information of Ref. \cite{PhysRevLett.108.190504} and it has taken into consideration of the storage efficiency of $6.9\%$ and the Poissonian distribution of input photons \cite{gundougan2015solid}. The storage fidelities for all input levels are far beyond the classical bound, indicating the quantum nature of this device.

\begin{table*}[h]
\caption{Fidelity of stored time-bin qubits}
\begin{ruledtabular}
\begin{tabular}{c|ccccc}
\diagbox{Photon Number}{Fidelity}{Input State} & $\ket{e}$ & $\ket{l}$ & $\ket{e}+i\ket{l}$ & $\ket{e}+e^{i3\pi/4}\ket{l}$ & $ F_T $ \\
\hline
0.2 & $98.4\%\pm0.2\%$ & $98.2\%\pm0.2\%$ & $96.4\%\pm0.6\%$ & $95.8\%\pm0.6\%$ & $96.8\%\pm0.3\%$ \\
0.4 & $98.8\%\pm0.1\%$ & $99.3\%\pm0.1\%$ & $97.5\%\pm0.3\%$ & $96.9\%\pm0.4\%$ & $97.8\%\pm0.2\%$ \\
0.8 & $99.1\%\pm0.1\%$ & $99.3\%\pm0.1\%$ & $97.8\%\pm0.2\%$ & $98.0\%\pm0.2\%$ & $98.3\%\pm0.1\%$ \\
1.6 & $99.4\%\pm0.04\%$ & $99.6\%\pm0.04\%$ & $97.9\%\pm0.2\%$ & $97.7\%\pm0.2\%$ & $98.3\%\pm0.1\%$ \\
3.2 & $99.6\%\pm0.03\%$ & $99.7\%\pm0.02\%$ & $98.0\%\pm0.1\%$ & $97.7\%\pm0.1\%$ & $98.5\%\pm0.04\%$ \\
\end{tabular}
\end{ruledtabular}
\end{table*}

\begin{figure}
%\begin{center}
\includegraphics [width= 0.8 \columnwidth]{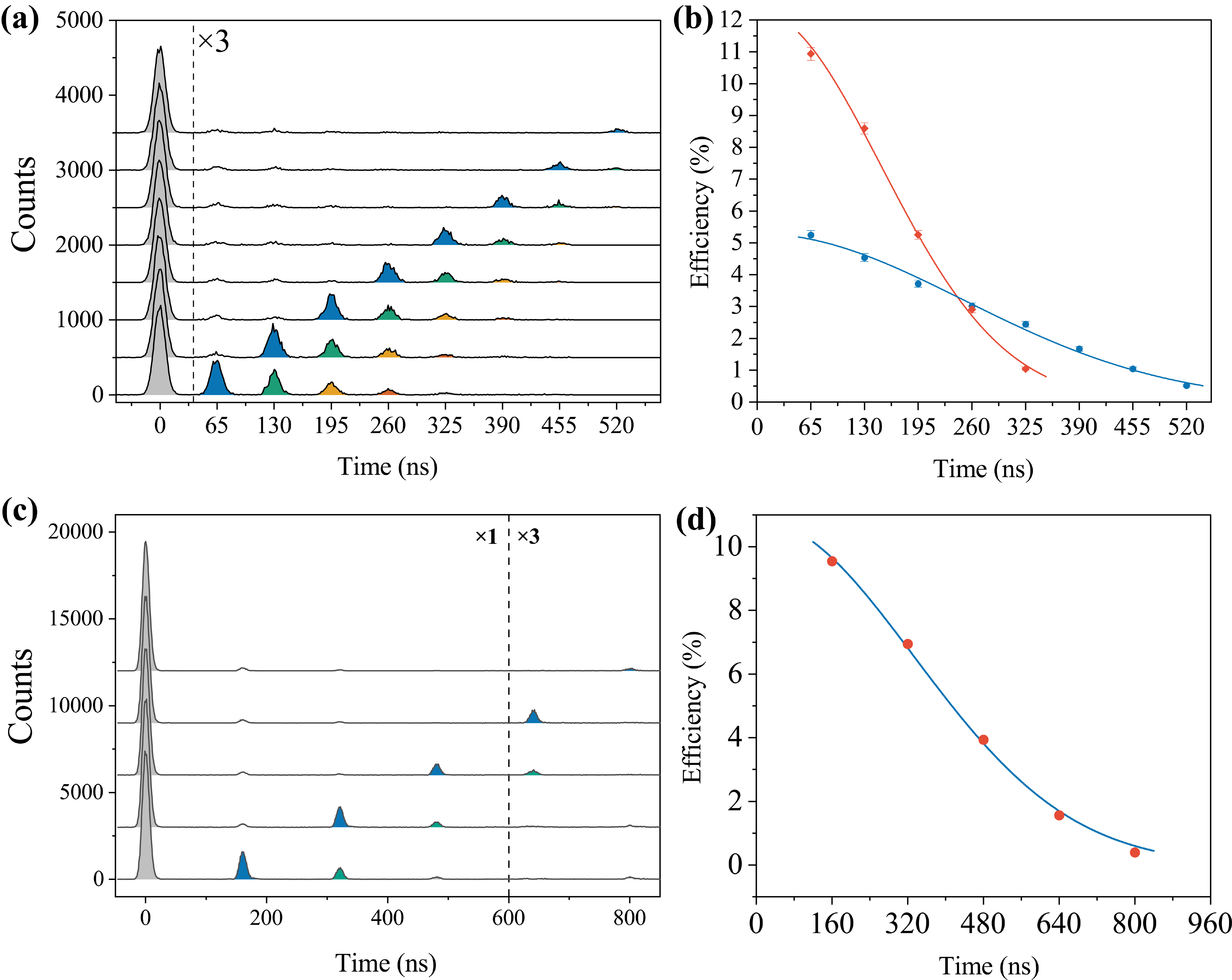}
\caption{\label{fig:echo} (a) Photon-counting histograms for on-demand storage with an input of $\mu=0.4$ using an AFC prepared with a finesse of $14.5\pm0.3$. 500 counts are shifted along y-axis between each set of data. (b) The storage efficiencies of two AFCs with different finesses. The red and blue data sets correspond to AFCs with finesses of $8.7\pm0.2$ and $14.5\pm0.3$, respectively. (c) Photon-counting histograms for on-demand storage with an input of $\mu=0.8$ using an AFC prepared with the $1/\Delta$ storage time of 160 ns and a finesse of $7.8\pm0.1$. 3000 counts are shifted along y-axis between each set of the data. (d) The storage efficiency as a function of the storage time of the AFC with $1/\Delta$ storage time of 160 ns. }
%\end{center}
\end{figure}

\begin{figure}
%\begin{center}
\includegraphics [width= 0.8 \columnwidth]{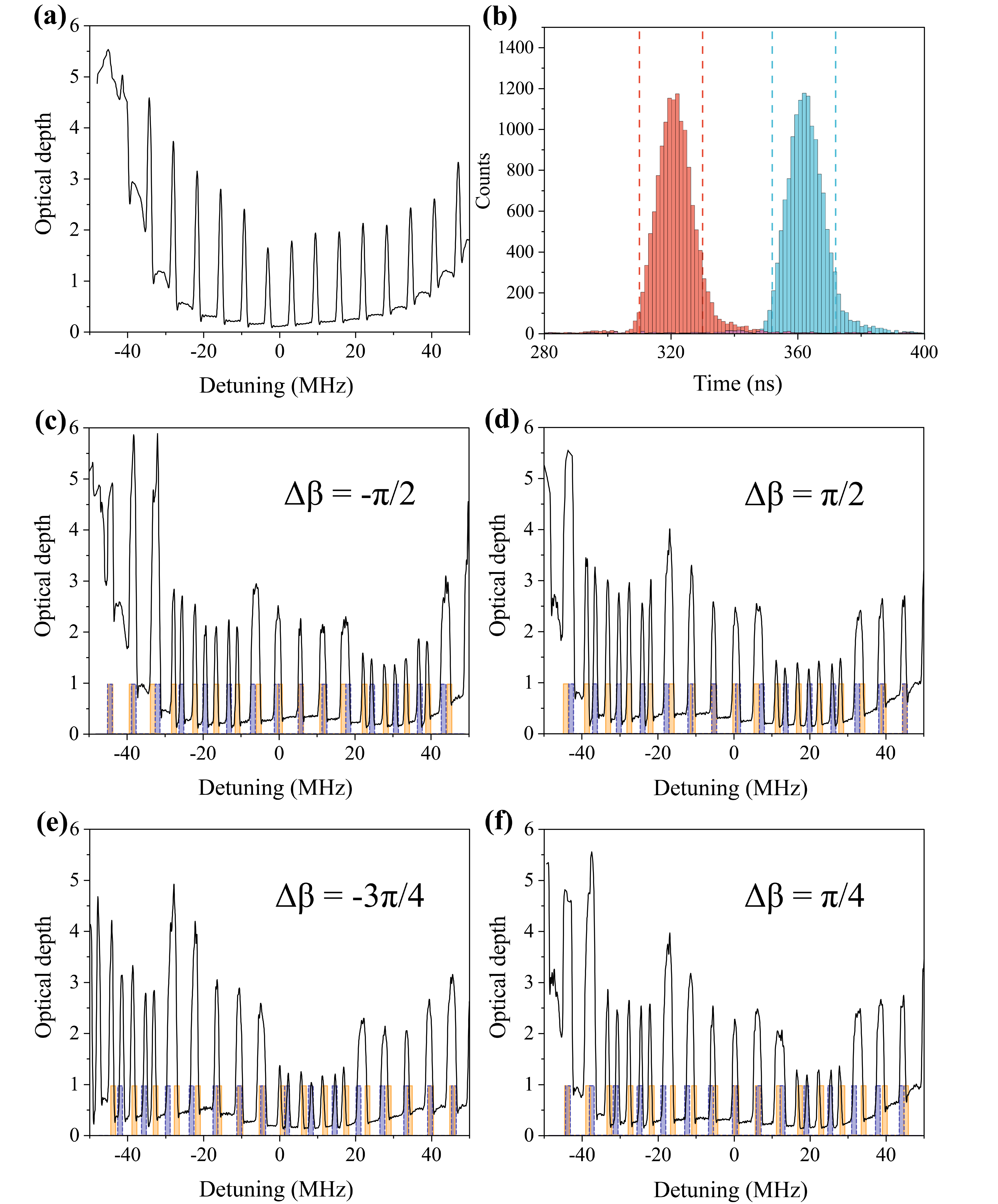}
\caption{\label{fig:afc} Demonstration of the AFC structures. (a) The AFC structure with a $1/\Delta$ storage time of 160 ns. (b) Photon-counting histograms for storage of time-bin states $\ket{e}$ (red) and $\ket{l}$ (blue) with $\mu=0.8$. The storage time of the AFC is $2/\Delta=$320 ns. The dashed lines indicate the integration windows (20 ns) for calculations of the fidelity. (c),(d),(e) and (f)  Double-AFC structures of various $\Delta\beta$. The yellow and blue strips are the simulated results of these double AFC structures. }
%\end{center}
\end{figure}

\begin{figure}
%\begin{center}
\includegraphics [width= 0.8 \columnwidth]{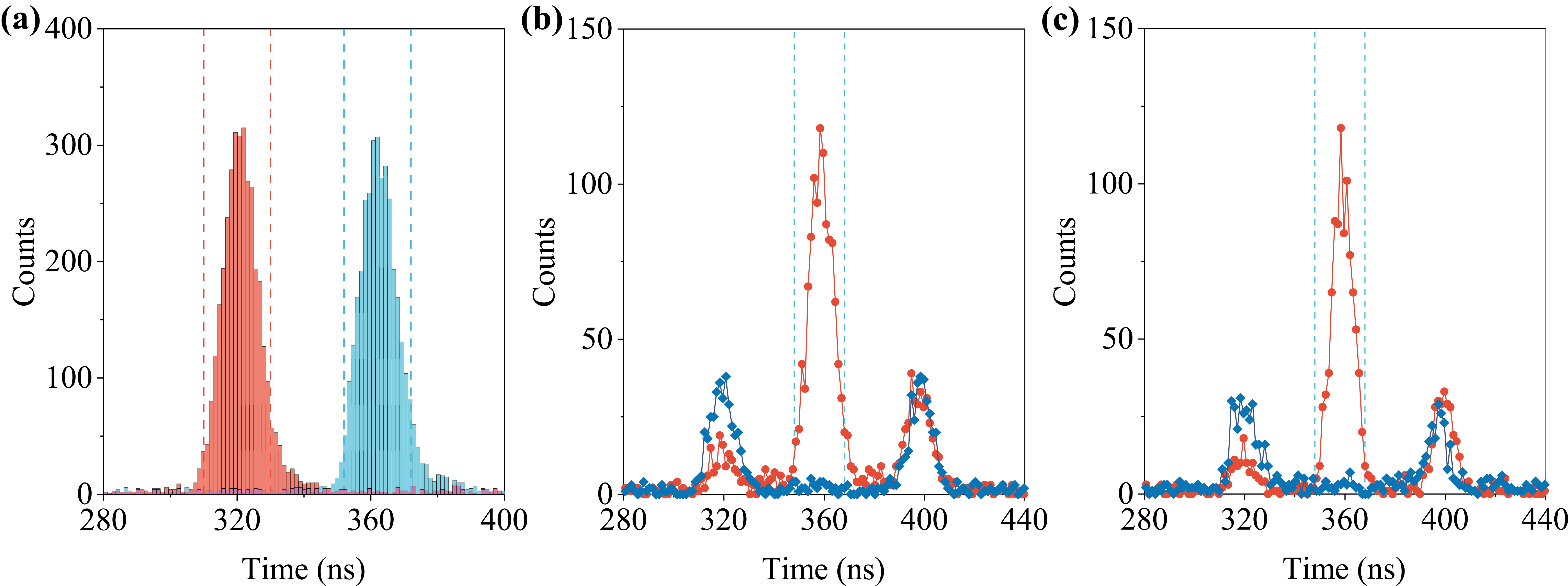}

\caption{\label{fig:el} Storage of time-bin qubits with $\mu=0.2$. (a) Photon-counting histograms for storage of inputs $\ket{e}$ (red) and $\ket{l}$ (blue). The storage time of the AFC is $2/\Delta=$320 ns. The fidelity is $98.4\%\pm0.2\%$ and $98.2\%\pm0.2\%$ for $\ket{e}$ and $\ket{l}$, respectively.  (b) and (c) present the photon-counting histograms for the constructive (shown in red) and destructive (shown in blue) interference for input states of $\ket{e}+i\ket{l}$ (\textbf{b}) and $\ket{e}+e^{i3\pi/4}\ket{l}$ (\textbf{c}), respectively. The fidelity is $96.4\%\pm0.6\%$ and $95.8\%\pm0.6\%$ for $\ket{e}+i\ket{l}$ and $\ket{e}+e^{i3\pi/4}\ket{l}$, respectively. The dashed lines indicate the integration windows (20 ns) for calculations of the fidelity.}
%\end{center}
\end{figure}

\end{document}